\documentclass[12pt,thmsa,titlepage]{article}
\usepackage{amssymb}

\input tcilatex

\begin{document}

\begin{center}
\textbf{METHOD OF LASER-RADIATION GUIDING IN PLASMA} Arsen G. Khachatryan 

\textit{Yerevan Physics Institute, Alikhanian Brothers Street 2, } \textit{%
Yerevan 375036, Armenia}
\end{center}

\begin{quotation}
The diffraction broadening of the intense laser radiation restricts its
efficient use in many applications. Proposed in the present work is a method
for laser radiation guiding in a density channel formed in plasma by a
relativistic electron beam. The conditions and parameters of the
relativistic beam ensuring the guiding by means of the proposed method have
been examined.
\end{quotation}

PACS number(s): 52.40.Mj, 52.40.Nk

The progress in the technology of high-intensity lasers opens new
opportunities for use of lasers in many branches of science and industry.
Last years the chirped-pulse amplification technique [1] permitted the
production of subpicosecond laser pulses of multiterrawatt power with peak
intensity up to 10$^{19}W/cm^2$ [2]. With intensities as such we practically
have to do with a new interaction range of laser radiation with matter,
where the role played by the nonlinear effects is often essential. At
present the interactions of high-power laser radiation with plasma are
actively investigated in connection with different applications: the
excitation of strong plasma wake waves for acceleration of charged particles
with acceleration rates to tens of $GeV/m$ [3]; generation, due to nonlinear
interaction with plasma, of radiation at harmonics of carrier laser
frequency [4]; the ``photon acceleration'' [5]; X-ray sources [6] etc. Note
also such application ranges of laser radiation as the Compton scattering,
laser cooling of charged particle beams, the inertial fusion.

The diffraction broadening of laser radiation is one of the principal
phenomena (and frequently the primary phenomenon) inhibiting the effective
use of the energy of laser in many applications. In vacuum, the laser spot
size $r_s$ grows with the longitudinal coordinate according to the formula $%
r_s=r_0(1+z^2/Z_R^2)^{1/2}$ , where $Z_R=\pi r_0^2/\lambda $ is the Rayleigh
length, $r_0$ is the minimum spot size at the focal point and $\lambda $ is
the laser wavelength. Owing to that, the intensity of radiation quickly
decreases as the laser beam propagates. For high-intensity laser pulses the
value of $Z_R$ is usually of the order of several millimeters. For instance,
in the Laser Wakefield Accelerator (LWFA) scheme the increment in the energy
of electrons accelerated by the longitudinal field of wake wave, excited by
a short laser pulse in plasma, is limited by the value $e\pi Z_RE_z$ [3],
where $E_z$ is the amplitude of accelerating electric field of the plasma
wave, $e$ is the charge of electron. Thus, without optical guiding the
diffraction limits the laser-matter interaction distance to a few Rayleigh
length.

In a medium, in particular, in plasma, if the index of refraction is maximum
at the axis of laser beam and decreases in the radial direction to its
periphery, one can eliminate or slow down the process of diffraction
broadening of laser radiation (see review in Ref. [7] and numerous
references therein). For the laser radiation with power $P>P_c=2c(e/r_e)^2[%
\omega /\omega _{pe}(r=0)]^2\approx 17[\omega /\omega _{pe}(r=0)]^2GW$,
where $r_e=e^2/m_ec^2$ is the classical electron radius, $\omega $ is the
frequency of laser radiation, $\omega _{pe}$ is the plasma frequency, there
takes place a relativistic self-focusing\textit{. }However, for short pulses
with the length $l\lesssim \pi c\,/\omega _{pe}$, the relativistic
self-focusing proves inefficient for prevention of diffraction broadening
[7]. In the experiments the plasma channel is usually formed in the gas or
plasma by a laser pulse that, in its turn, is also subject to diffraction.
E.g., the parabolic plasma density profile $n_p=n_0+\Delta nr^2/r_0^2$ may
provide the guiding of low-intensity [$a_0^2=(eE_0/m_ec\omega )^2\ll 1$,
where $E_0$ is the amplitude of laser radiation] Gaussian laser beam if $%
\Delta n\geq \Delta n_c=1/\pi r_er_0^2=1.13\times 10^{20}/r_0^2[\mu m]$ $%
cm^{-3}$ [7]. The guiding of laser radiation in the preformed plasma density
channel at distances from several millimeters to 1-3 $cm$ has been
demonstrated by different groups of researchers. In the present work for
formation of a plasma channel we propose to use a long relativistic electron
beam (REB). REB\ may traverse, without any essential changes in parameters,
the distances in plasma that are much longer than the Rayleigh length of
high-intensity laser pulses. The main advantage of the proposed method
consists, hence, in the fact that REB can form a plasma channel with lengths
much exceeding those obtained in the experiments under survey. So, the
method in question could provide longer-term interaction of high-intensity
laser radiation with plasma, as well as with relativistic electrons.

Consider the propagation of a cylindrical electron beam with velocity $%
\mathbf{v}_0=\mathbf{e}_zv_0$ in cold homogeneous plasma. From the Poisson
equation, the equation of motion and the continuity equation for plasma
electrons (the plasma ions are taken to be immobile due to their large mass)
we have:

\begin{equation}
\frac{\partial ^2\delta n_e}{\partial t^2}+\omega _{pe}^2(\delta n_e+n_b)=0,
\tag{1}
\end{equation}
where $\delta n_e=n_e-n_0$, $\,n_e$ is the density of plasma electrons, $n_0$
is their equilibrium density, $n_b$ is the density of electrons in the beam, 
$\omega _{pe}=(4\pi n_0e^2/m_e)^{1/2}$ is the plasma frequency. In case of
long beam with the length much exceeding the plasma wavelength $\lambda
_p=2\pi v_0/\omega _{pe}$, and density $\,n_b=n_b(r)$, one can omit the
first term in Eq. (1). Then one has

\begin{equation}
n_e(r)=n_0-n_b(r).  \tag{2}
\end{equation}
The plasma electrons are blown out of the beam and a density profile (2) is
established. Though Eq. (1) was obtained for the linear case when $n_b\ll
n_0 $, the expression (2) holds true also for $n_b\lesssim n_0$. So, for a
function $n_b(r)$ decreasing as $r$ we have a plasma electron density
channel. At the same time, the summary density of electrons is constant, $%
n_e+n_b=n_0=const$, and is equal to the density of ions, and, therefore, the
force acting on the ions of plasma is zero. In this paper we shall show that
in spite of the summary density of electrons during the traversal of long
electron beam through plasma is constant, the guiding of laser radiation in
this case is possible. The problem of the formation of plasma channel and
the stability of electron beam will be discussed below.

For consideration of the problem of laser radiation guiding in the plasma
density channel with density (2), first examine the dispersion properties of
electromagnetic (EM) waves in the channel. From the Maxwell equations one
can has for the electric field strength of EM wave

\begin{equation}
rotrot\mathbf{E}=\mathbf{\nabla (\nabla E)-}\Delta \mathbf{E=}%
-c^{-2}(\partial ^2\mathbf{E/}\partial t^2+4\pi \partial \mathbf{j}/\partial
t),  \tag{3}
\end{equation}
where $\mathbf{j}=-e(n_e\mathbf{v}_e+n_b\mathbf{v}_b)$ is the density of
current, $\mathbf{v}_e$, $\mathbf{v}_b$ are the velocities of plasma
electrons and beam respectively. For a linearly polarized wave, $\mathbf{E}=%
\mathbf{e}_xE_x=\mathbf{e}_xE_0\exp [i(\omega t-kz)]$, $\mathbf{B}=\mathbf{e}%
_y(ck/\omega )E_x$, where $\mathbf{B}$ is the vector of magnetic induction,
we obtain from (3)

\begin{equation}
(k^2c^2-\omega ^2)E_x\mathbf{=}4\pi e[n_e(r)\partial v_{ex}\mathbf{/}%
\partial t+n_b(r)\partial v_{bx}\mathbf{/}\partial t].  \tag{4}
\end{equation}
On obtaining the expression (4) we assumed that the response of plasma
channel to the propagation of the laser radiation is linear, hence the
change of $n_e$ and $n_b$ under action of EM wave is negligible (that is,
nonlinear terms $v_{ex}\partial n_e\mathbf{/}\partial t$ and $v_{bx}\partial
n_b\mathbf{/}\partial t$ are threw away). This takes place when $%
a_0^2=(eE_0/m_ec\omega )^2\ll 1$ [3]. To be exact, in our case Eq. (4) is
valid if the change in the density of plasma electrons under the action of
EM wave, that is proportional to $a_0^2$, is much less than the change of
density under the action of electron beam, i. e. , $a_0^2\ll n_b/n_0$. The
derivatives of velocities $v_{ex}$ and $v_{bx}$ in the right hand side of
Eq. (4) will be obtained from the equations of motion:

\begin{equation}
\frac{\partial v_{ex}}{\partial t}=-\frac e{m_e}E_x,  \tag{5.1}
\end{equation}

\begin{equation}
(\frac \partial {\partial t}+v_{bz}\frac \partial {\partial z})(v_{bx}\gamma
_b)=-\frac e{m_e}(E_x+\frac{v_{bz}}cB_y),  \tag{5.2}
\end{equation}
where $\gamma _b=(1-\mathbf{v}_b^2/c^2)^{-1/2}$ $\,$is a relativistic
factor. In case of relativistic electron beam, when $\gamma
_0=(1-v_0^2/c^2)^{-1/2}\gg 1$, we can put $v_{bx}\ll v_{bz}\approx
v_0\approx c$. Taking into account that $\gamma
_0^2v_{bx}^2/c^2=(eE_x/m_ec\omega )^2=a^2\ll 1$, we obtain from (5.2) $%
\partial v_{bx}/\partial t=-(e/m_e\gamma _0)E_x$. Substituting this
expression and Eq. (5.1) into Eq. (4) and taking into account (2) we obtain
the following dispersion relation

\begin{equation}
\omega ^2=k^2c^2+\omega _{pe}^2[1-\alpha (1-\gamma _0^{-1})]\mathbf{\approx }%
k^2c^2+\omega _{pe}^2(1-\alpha ),  \tag{6}
\end{equation}
where $\alpha =n_b(r)/n_0$. In the absence of beam ($\alpha =0$) one has
from (6) an ordinary dispersion relation for transverse waves in cold
homogeneous plasma. The expression (6) is valid also for the case of a
circularly polarized wave, because one can write that as a superposition of
two linearly polarized waves. So, although the summary density of electrons $%
n_b+n_e=const,$ EM wave does not ''feel'' the relativistic electron beam
owing to the fact that the relativistic mass of beam electrons is much
larger than the mass of plasma electrons, just as the ions make negligible
contribution to the dispersion relation thanks to their large mass. For this
reason, instead of REB one can use a beam of relativistic or nonrelativistic
negatively charged ions. It is noteworthy also that instead of a continuous
beam one can use a long succession of bunches, the separation between which
is much less than the plasma wavelength. EM wave can propagate both along
and opposite REB. The latter is important for such applications as Compton
scattering or a plasma-based free electron laser [8].

From Eq. (6) we have for the index of refraction $N=ck/\omega $ of EM\ wave
(laser radiation)

\begin{equation}
N^2=1-[1-\alpha (r)]\omega _{pe}^2/\omega ^2.  \tag{7}
\end{equation}
The guiding of laser radiation is possible when the condition $dN/dr=(\omega
_{pe}^2/2N\omega ^2)d\alpha /dr<0$ is observed. Consider now a parabolic
profile of the electron beam density, $\alpha =\alpha _0(1-r^2/r_b^2)$, $%
r<r_b$, and the Gaussian profile of the laser radiation, $%
\,a=(a_0r_0/r_s)\exp (-r^2/r_s^2)$.\thinspace \thinspace Then $%
n_e(r)=n_e(0)+n_b(0)r^2/r_b^2$, where $n_e(0)=n_0-n_b(0)$. In this case the
plasma electron density channel produced by REB provides the laser radiation
guiding when (see Chapter VI in Ref. [7])

\begin{equation}
n_b(0)>\Delta n_c(r_br_0)^2/r_s^4.  \tag{8}
\end{equation}
For instance, when $r_b=r_s=2r_0$ and $r_0=500\mu m$,we have $%
n_b(0)>1.13\times 10^{14}cm^{-3}$, and the Rayleigh length is $Z_R\approx
7.85cm$ for the wavelength of laser radiation $\lambda =10\mu m$.

The linear case ($\delta n_e\ll n_0$) considered above was when $a_0^2\ll
\alpha _0\ll 1$. At the violation of this condition the mathematical
description of the problem is complicated, but the guiding again is
possible. Moreover, one can weaken the condition of guidance. In case when $%
a_0^2\gtrsim \alpha _0$, one can weaken the condition of guidance (8) for
the long laser beam ($l\gg \lambda _p$), first, because of the fact that the
plasma electrons are blown out of the channel by the laser radiation owing
to which the effect of guiding is amplified, and, second, due to the effect
of relativistic self-focusing. When $a_0^2\ll 1$, the condition of laser
radiation guiding in this case takes on the form (see Ref. [7]):

\[
\frac P{P_c}+\frac{a_0^2}2\frac{r_0^2}{r_s^2}+\frac{n_b(0)}{\Delta n_c}\frac{%
r_s^4}{(r_br_0)^2}>1. 
\]
The large gradient of plasma density in the channel may be formed by a dense
REB when $n_b\gtrsim n_0$. In this case there is formed a range close to the
beam axis, where all plasma electrons are driven out. For a thin beam ( with
the radius $r_b\ll \lambda _p$) such a range is formed when $n_b>n_0$, and
for a broad beam ($r_b\gg \lambda _p$) - when $n_b>n_0/2$, the diameter of
range increasing with the beam density [9].

Now consider the problems of plasma channel formation and the plasma
stability. To avoid the excitation of plasma wave by the leading edge of
electron beam, the density of REB at the entry into plasma should grow
rather slowly, to wit as $t_b\gg \omega _{pe}^{-1}$ where\textit{\ }$t_b$ is
the rise time of beam density. However, for applications connected with the
acceleration of charged particles the excitation of plasma wave by the
electron beam may prove desirable. During the flight of REB through plasma
it is subject to different forms of instability. The condition of neglect of
an instability may be written in the form $\delta \Delta t\lesssim 1$, where 
$\delta $ is an increment of the instability and $\Delta t$ \thinspace is
the time of flight of electron beam through plasma; for relativistic beam $%
\Delta t\approx l\,/c$, where $l$ is the length of plasma column. The most
quickly developing instability is the beam-plasma instability as a result of
which a plasma wave with wave number $k\approx \omega _{pe}/v_0$ is excited.
The development of beam-plasma instability leads eventually to a breakage of
beam into bunches flying with the period of plasma wave. The increment of
beam-plasma instability for waves propagating along the direction of
electron beam flight (i.e., along $z$ axis) is (see, e.g., Ref. [10])

\begin{equation}
\delta =(3^{1/2}/2^{4/3})\omega _{pe}(n_b/n_0)^{1/3}/\gamma _0.  \tag{9}
\end{equation}
It follows from (9) that the increment of beam-plasma instability decreases
with increasing energy of beam electrons, i.e., for sufficiently large $%
\gamma _0$ one can neglect the instability. According to (9), one can
neglect the instability for

\[
\gamma _0\gtrsim 1.3\times 10^{-6}n_0^{1/2}[cm^{-3}](n_b/n_0)^{1/3}l[cm]. 
\]
For $n_0=10^{14}cm^{-3}$, $n_b/n_0=0.1$ and $l=50cm$, one has $\gamma
_0\gtrsim 300$. The development of beam-plasma instability may be desirable
for a number of applications, e.g., for generation of picosecond and
femtosecond electron bunches. In this case the wake wave excited by a short
laser pulse in the plasma channel produced by the electron beam would enable
one to control the beam instability.

Consider the problem of guiding from an energetic point of view. The energy
density of REB and laser beam are respectively $W_{REB}=m_ec^2\gamma _0n_b$
and $W_{Las}=E_0^2/8\pi =(m_ec\omega /e)^2a_0^2/8\pi $, and for their ratio
one has

\begin{equation}
\kappa \equiv \frac{W_{REB}}{W_{Las}}=\frac{2\gamma _0(n_b/n_0)}{a_0^2}%
\left( \frac \lambda {\lambda _p}\right) ^2.  \tag{10}
\end{equation}
The wavelength $\lambda $ of modern high-intensity lasers is in order of
several micrometer (the value of $\lambda \approx 1\mu m$ is typical for a
solid state laser and $\lambda \approx 10\mu m$ for the CO$_{\text{2}}$
laser). In the case $n_0=10^{14}cm^{-3}$, $n_b/n_0=0.1$, $\gamma _0=300$, $%
\lambda =1\mu m$, and $a_0^2=0.05$ (that corresponds to the circularly
polarized laser beam intensity $I\approx 1.4\times 10^{17}W/cm^2$) from
expression (10) follows $\kappa \approx 1.08\times 10^{-4}$. When $%
n_0=10^{16}cm^{-3}$, $n_b/n_0=0.1$, $\gamma _0=100$, $\lambda =10\mu m$, and 
$a_0^2=0.1$ ($I\approx 2.8\times 10^{15}W/cm^2$) one has $\kappa \approx
0.18 $. Thus, REB allows to guide laser radiation with the energy density
much exceeding that of REB. The total energy of REB $\varepsilon _{REB}\sim
\pi r_b^2l_{REB}W_{REB}$, for practically interesting parameters of the
problem, can be more or less than the total energy of laser beam $%
\varepsilon _{REB}\sim \pi r_0^2l_{Las}W_{Las}$; here $l_{REB}$ and $l_{Las}$
are respectively the length of REB and laser beam.

It was shown above that the relativistic electron beam can form a channel in
plasma, in which the laser radiation guiding is possible. The proposed
guiding method is based on the capacity of relativistic electron beam to
traverse, without essential change in parameters, the distances in plasma
much larger than the diffraction length of high-intensity laser radiation.
The plasma electrons are blown out of the range occupied by REB, as a result
of which a plasma channel is formed in plasma, the density in which
increases in radial direction. It was shown that owing to large relativistic
factor of REB, its contribution to the dispersion properties of channel is
negligible. The method under consideration permits an essential increase in
the interaction time of laser radiation with plasma and electron beam and,
hence, in the efficiency of using the radiation energy.

The author gratefully acknowledges helpful discussions with Dr. B. Hafizi,
Dr. R. Hubbard, and Dr. P. Sprangle (Naval Research Laboratory, Washington,
DC).

\begin{center}
\textbf{REFERENCES}
\end{center}

[1] \thinspace D. Strickland and G. Mourou, Opt. Commun., \textbf{56}, 219
(1985).

[2] M. D. Perry, In: \textit{Advanced accelerator concepts, }edited by P.
Schoessow, AIP Conference Proceedings No. 335, 1995.

[3] E. Esarey, P. Sprangle, J. Krall, and A. Ting, IEEE Trans. Plasma Sci. 
\textbf{24}, 252 (1996).

[4] P. Sprangle, E. Esarey, and A. Ting, Phys. Rev. A \textbf{41}, 4463
(1990).

[5] S. C. Wilks, J. M. Dawson, W. B. Mori, T. Katsouleas, and M. E. Jones,
Phys. Rev. Lett. \textbf{62}, 2600 (1989).

[6] \thinspace N. H. Burnett and P. B. Corkum, J. Opt. Soc. Am. B,\textbf{\ 6%
}, 1195 (1989).

[7] E. Esarey, P. Sprangle, J. Krall, and A. Ting, IEEE J. Quantum
Electron., \textbf{33}, 1879 (1997).

[8] K. Nakajima, M. Kando, T. Kawakubo, T. Nakanishi, and A. Ogata, Nucl.
Instr. and Meth. A \textbf{375}, 575 (1996).

[9] I. A. Kotel'nikov and V. N. Khudik, Fizika Plazmy, \textbf{23,} 146
(1997) [Plasma Phys. Rep., \textbf{23}, 130 (1997)].

[10] A. B. Mikhailovskii, \textit{Theory of Plasma Instabilities}
(Consultants Bureau, New York, 1974), Vol. I.

\end{document}